\documentclass[prd, twocolumn, nofootinbib]{revtex4-2}

\usepackage{graphicx}
\graphicspath{ {./images/} }

\usepackage{amsmath}
\usepackage{amssymb}

\usepackage{enumerate}
\usepackage{empheq}

\usepackage{natbib}

\DeclareUnicodeCharacter{2009}{\,} 
\usepackage{listings}
\usepackage{multirow}
\usepackage{comment}

\usepackage[usenames,dvipsnames]{xcolor}
\usepackage[hyperfootnotes=false]{hyperref}
\hypersetup{
  colorlinks,
  citecolor=blue,
  linkcolor=blue,
  urlcolor=blue,
}
  
\newcommand{\lb}{\left(}
\newcommand{\rb}{\right)}

\newcommand{\lc}{\left\lbrace}
\newcommand{\rc}{\right\rbrace}

\newcommand{\tides}{\Lambda}                        
\newcommand{\spin}{\chi}                            
\newcommand{\chirp}{\mathcal{M}}                    
\newcommand{\smr}{\eta}                             
\newcommand{\chieff}{\chi_\mathrm{eff}}         
\newcommand{\chia}{\chi_\mathrm{a}}         
\newcommand{\lameff}{\Tilde{\tides}}                
\newcommand{\dlam}{\delta\tides}                    

\newcommand{\object}{\mathcal{O}}                   
\newcommand{\bh}{\text{BH}}                         
\newcommand{\ns}{\text{NS}}                         
\newcommand{\snr}{\rho}                             
\newcommand{\angproj}{\Theta}                       
\newcommand{\obs}{_{\text{obs}} }                   

\newcommand{\detec}{\text{det}}                     

\newcommand{\pmf}{P}                                
\newcommand{\pdf}{p}                                

\newcommand{\numdens}{\mathcal{N}}                  
\newcommand{\rate}{\mathcal{R}}                     
\newcommand{\parvec}{\mathbf{\theta}}               
\newcommand{\pnpar}{\parvec}                        
\newcommand{\model}{\lambda}                        

\newcommand{\chims}{\chi_{\text{ms}}}               

\newcommand{\data}{d}                          

\newcommand{\result}[1]{#1}

\newcommand{\AFzeroCone}{\result{\ensuremath{0.49^{+0.18}_{-0.49}}}}
\newcommand{\AFzeroCtwo}{\result{\ensuremath{0.46^{+0.17}_{-0.46}}}}
\newcommand{\AFzeroCthree}{\result{\ensuremath{0.38^{+0.12}_{-0.38}}}}
\newcommand{\AFzeroCfour}{\result{\ensuremath{0.33^{+0.09}_{-0.33}}}}
\newcommand{\AFzeroCfive}{\result{\ensuremath{0.30^{+0.08}_{-0.30}}}}

\newcommand{\AGzeroCone}{\result{\ensuremath{0.49^{+0.18}_{-0.49}}}}
\newcommand{\AGoneCone}{\result{\ensuremath{0.50^{+0.46}_{-0.22}}}}
\newcommand{\AGzeroCtwo}{\result{\ensuremath{0.47^{+0.17}_{-0.47}}}}
\newcommand{\AGoneCtwo}{\result{\ensuremath{0.49^{+0.18}_{-0.49}}}}
\newcommand{\AGzeroCthree}{\result{\ensuremath{0.42^{+0.14}_{-0.42}}}}
\newcommand{\AGoneCthree}{\result{\ensuremath{0.49^{+0.18}_{-0.48}}}}
\newcommand{\AGzeroCfour}{\result{\ensuremath{0.35^{+0.10}_{-0.35}}}}
\newcommand{\AGoneCfour}{\result{\ensuremath{0.48^{+0.17}_{-0.48}}}}
\newcommand{\AGzeroCfive}{\result{\ensuremath{0.33^{+0.09}_{-0.33}}}}
\newcommand{\AGoneCfive}{\result{\ensuremath{0.48^{+0.18}_{-0.48}}}}

\newcommand{\BFzeroCone}{\result{\ensuremath{0.50^{+0.49}_{-0.18}}}}
\newcommand{\BFzeroCtwo}{\result{\ensuremath{0.54^{+0.46}_{-0.16}}}}
\newcommand{\BFzeroCthree}{\result{\ensuremath{0.53^{+0.45}_{-0.18}}}}
\newcommand{\BFzeroCfour}{\result{\ensuremath{0.52^{+0.37}_{-0.25}}}}
\newcommand{\BFzeroCfive}{\result{\ensuremath{0.52^{+0.35}_{-0.27}}}}
\newcommand{\BGzeroCone}{\result{\ensuremath{0.50^{+0.44}_{-0.24}}}}
\newcommand{\BGoneCone}{\result{\ensuremath{0.51^{+0.48}_{-0.20}}}}
\newcommand{\BGzeroCtwo}{\result{\ensuremath{0.49^{+0.18}_{-0.49}}}}
\newcommand{\BGoneCtwo}{\result{\ensuremath{0.50^{+0.50}_{-0.18}}}}
\newcommand{\BGzeroCthree}{\result{\ensuremath{0.48^{+0.17}_{-0.48}}}}
\newcommand{\BGoneCthree}{\result{\ensuremath{0.51^{+0.49}_{-0.18}}}}
\newcommand{\BGzeroCfour}{\result{\ensuremath{0.47^{+0.16}_{-0.47}}}}
\newcommand{\BGoneCfour}{\result{\ensuremath{0.51^{+0.48}_{-0.19}}}}
\newcommand{\BGzeroCfive}{\result{\ensuremath{0.47^{+0.19}_{-0.42}}}}
\newcommand{\BGoneCfive}{\result{\ensuremath{0.50^{+0.49}_{-0.19}}}}

\newcommand{\CFzeroCone}{\result{\ensuremath{0.54^{+0.46}_{-0.17}}}}
\newcommand{\CFzeroCtwo}{\result{\ensuremath{0.60^{+0.40}_{-0.13}}}}
\newcommand{\CFzeroCthree}{\result{\ensuremath{0.69^{+0.31}_{-0.08}}}}
\newcommand{\CFzeroCfour}{\result{\ensuremath{0.75^{+0.25}_{-0.05}}}}
\newcommand{\CFzeroCfive}{\result{\ensuremath{0.77^{+0.23}_{-0.05}}}}
\newcommand{\CGzeroCone}{\result{\ensuremath{0.53^{+0.47}_{-0.17}}}}
\newcommand{\CGoneCone}{\result{\ensuremath{0.50^{+0.47}_{-0.21}}}}
\newcommand{\CGzeroCtwo}{\result{\ensuremath{0.57^{+0.43}_{-0.15}}}}
\newcommand{\CGoneCtwo}{\result{\ensuremath{0.51^{+0.49}_{-0.18}}}}
\newcommand{\CGzeroCthree}{\result{\ensuremath{0.63^{+0.37}_{-0.12}}}}
\newcommand{\CGoneCthree}{\result{\ensuremath{0.52^{+0.48}_{-0.17}}}}
\newcommand{\CGzeroCfour}{\result{\ensuremath{0.68^{+0.32}_{-0.08}}}}
\newcommand{\CGoneCfour}{\result{\ensuremath{0.54^{+0.46}_{-0.17}}}}
\newcommand{\CGzeroCfive}{\result{\ensuremath{0.72^{+0.28}_{-0.07}}}}
\newcommand{\CGoneCfive}{\result{\ensuremath{0.54^{+0.46}_{-0.17}}}}

\begin{document}

\title{Distinguishing between Black Holes and Neutron Stars within a Population of Weak Tidal Measurements}

\author{Michael M\"{u}ller}
\email{michael.mueller@uni-greifswald.de}
\affiliation{Institute of Physics, University of Greifswald, D-17489 Greifswald, Germany}

\author{Reed Essick}
\email{essick@cita.utoronto.ca}
\affiliation{Canadian Institute for Theoretical Astrophysics, University of Toronto, 60 St. George Street, Toronto, ON M5S 3H8}
\affiliation{Department of Physics, University of Toronto, 60 St. George Street, Toronto, ON M5S 1A7}
\affiliation{David A. Dunlap Department of Astronomy, University of Toronto, 50 St. George Street, Toronto, ON M5S 3H4}

\begin{abstract}
    We study the ability of tidal signatures within the inspiral of compact binaries observed through gravitational waves (GWs) to distinguish between neutron stars (NSs) and black holes (BHs).
    After quantifying how hard this measurement is on a single-event basis, we investigate the ability of a large catalog of GW detections to constrain the fraction of NS in the population as a function of mass: $f_\ns(m)$.
    Using simulated catalogs with realistic measurement uncertainty, we find that \result{$> O(200)$} events will be needed before we can precisely measure $f_\ns$, and catalogs of \result{$> O(100)$} events will be needed before we can even rule out the possibility that all low-mass objects are BHs with GW data alone (i.e., without electromagnetic counterparts).
    Therefore, this is unlikely to occur with advanced detectors, even at design sensitivity.
    Nevertheless, it could be feasible with next-generation facilities like Cosmic Explorer and Einstein Telescope.
\end{abstract}

\maketitle


\section{Introduction}
\label{sec:intro}

Tidal signatures are difficult to detect in gravitational waves (GWs) from individual compact binary coalescences (CBCs).
This poses a significant challenge if one is to understand the true nature of a specific event in terms of its matter content; that is, whether the binary contained a neutron star (NS) or only contained black holes (BHs).
Typically, the leading-order tidal signatures, often taken to be a smoking-gun signature of matter,\footnote{There is interest in other signatures, but these are often harder to detect. See, for example,~\citet{GW170817-pg} and~\citet{GW170817-remnant}.} become increasingly difficult to measure when the component masses are large because larger masses lead to more compact stars and correspondingly smaller tidal responses.
This makes direct measurements of the maximum mass of non-spinning NSs ($M_\mathrm{TOV}$) through tidal effects during GW inspirals impractical for individual events, since the difference between the tidal response of NSs at $M_\mathrm{TOV}$ and the BH limit of zero can be more than an order of magnitude smaller than the measurement uncertainty~\cite{chenDistinguishingHighmassBinary2020}.

However, this does not mean that we cannot extract information from a catalog of noisy tidal observations.
We use hierarchical Bayesian inference to combine many noisy observations of tidal signatures within a catalog of CBCs observed through GWs.
In our analysis, we make the simplifying assumptions that the overall mass distribution (comprising both NSs and BHs) and the equation of state (EoS) are known.
We then measure the fraction of the compact object population that is made up of NSs as a function of mass and determine how well this can be measured at different mass scales.
In reality, both the EoS and the shape of the overall population will not be perfectly known, and therefore our estimates are likely optimistic as they do not account for this additional uncertainty.

Current GW catalogs~\cite{GWTC-1, GWTC-2, GWTC-2.1, GWTC-3, GWTC-4} suggest that there is a sharp decrease in the number of compact objects within merging binaries above $\sim 2$-$3\,M_\odot$ \cite{abbottGWTC1GravitationalWaveTransient2019,abbottPopulationPropertiesCompact2021,farahBridgingGapCategorizing2021}.\footnote{The details of the inferred mass and spin distributions have changed somewhat with the addition of new events in GWTC-4.0~\cite{LIGOScientific:2025pvj}. However, the general trends observed previously are still present.}
Intriguingly, this is consistent with current estimates of $M_\mathrm{TOV}$ and begs the question of whether the drop-off occurs exactly at $M_\mathrm{TOV}$ or whether there are other factors governing the formation of such binaries that produce this effect.
Additionally, it is commonly assumed that all objects with masses $\leq M_\mathrm{TOV}$ are NSs, even without any clear tidal signature or electromagnetic counterpart.
We take a first step towards relaxing these assumptions and show how large GW catalogs must be before we can test these assumptions.\footnote{Other authors have also looked at the distinguishability between binary NS systems and NS-BH systems using tidal information. See, for example,~\citet{chenDistinguishingBinaryNeutron2020}.}

We consider simulated CBC signals with single-event measurement uncertainties consistent with expectations for the current ground-based GW interferometers: advanced LIGO~\cite{LIGO} and Virgo~\cite{Virgo}.
We do not consider KAGRA~\cite{KAGRA} or proposed next-generation detectors (Cosmic Explorer~\cite{CE} and Einstein Telescope~\cite{ET}), although we expect our conclusions to be similar for all detector networks.
This is because catalogs of low-mass events will be flux-limited, event with next-generation detectors~\cite{essickWhenSweatSmall2024}, and the tidal information for individual events will be similar between current and next-generation detectors (at a fixed signal-to-noise ratio).
As such, we expect similar constraints as a function of the number of detections.

Briefly, we confirm that it is more difficult to distinguish between NSs and BHs using adiabatic tidal measurements at high masses than at low masses~\cite{Crescimbeni:2024cwh, Golomb:2024mmt}.
We also find that precise constraints on the fraction of objects (with a particular mass) that are NSs will likely require catalogs of \result{$> O(200)$} events, even for low masses.
Nevertheless, we may still be able to answer some astrophysically relevant questions with as few as \result{$O(100)$} detections.

This work is structured as follows.
Sec.~\ref{sec:tidal signatures} discusses the basics of adiabatic tidal signatures within GW waveforms.
In Sec.~\ref{sec:sing_ev}, we employ Fisher matrix estimates to understand how the constraints on tides depend on the component masses.
This provides a qualitative understanding of how the population inference will be affected by different types of sources.
In Sec.~\ref{sec:population}, we implement a full hierarchical inference with mock data and a mixture model to account for the two
different species of compact objects (BHs and NSs).
We conclude in Sec.~\ref{sec:discussion}.


\section{Tidal signatures in CBCs}
\label{sec:tidal signatures}

\begin{figure}
    \centering
    \includegraphics[width=0.5\textwidth]{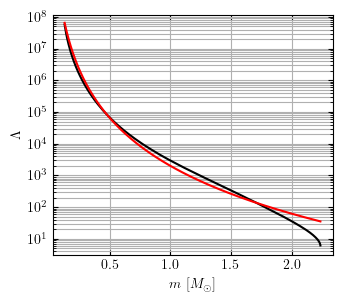}
    \caption{
        Tidal deformability ($\Lambda$) as a function of the gravitational mass ($m$).
        We show the (\emph{black}) maximum-likelihood EoS from \citet{essickAstrophysicalConstraintsSymmetry2021, essickDetailedExaminationAstrophysical2021a} and a (\emph{red}) simple power-law with $\Lambda \propto m^{-5}$.
    }
    \label{fig:ns_tides}
\end{figure}

\begin{figure*}
    \begin{center}
    \includegraphics[width=1.0\textwidth, clip=True, trim=0.0cm 1.2cm 0.0cm 0.75cm]{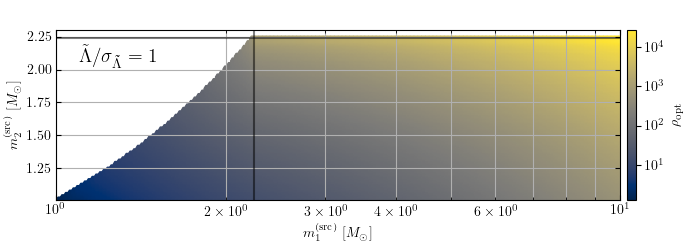}
    \includegraphics[width=1.0\textwidth, clip=True, trim=0.0cm 0.0cm 0.0cm 0.75cm]{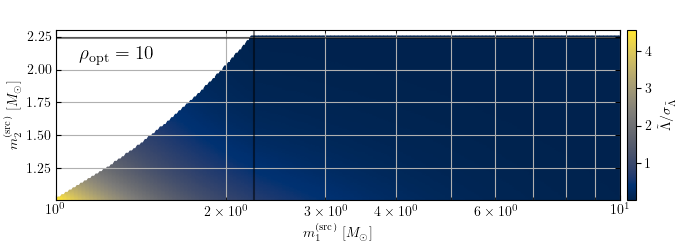}
    \end{center}
    \caption{
        Measurement uncertainty for $\lameff$ for individual events.
        (\emph{top}) $\rho_\mathrm{opt}$ required to achieve \result{$\lameff = \sigma_{\lameff}$} as a function of source-frame masses for our reference EoS (Fig.~\ref{fig:ns_tides}).
        (\emph{bottom}) Relative uncertainty ($\lameff / \sigma_{\lameff}$) at $\rho_\mathrm{opt}=10$.
        The boundary between NSs and BHs ($M_\mathrm{TOV}$) is denoted by black lines.
    }
    \label{fig:fish_tides_fixed_rel_uncert}
\end{figure*}

The leading-order tidal signature within CBC waveforms enters the phase at 5th post-Newtonian (PN) order as a linear combination of the tidal deformabilities of the binary's components weighted by their masses

\begin{equation} \label{eqn:lameff}
    \lameff = \left(\frac{16}{13}\right) \frac{(m_1+12m_2)m_1^4 \Lambda_1 + (m_2 + 12m_1)m_2^4 \Lambda_2}{\lb m_1 + m_2\rb^5}
\end{equation}
where $m_{1,2}$ are the masses of the binary's components and $\Lambda_{1,2}$ are their dimensionless tidal deformabilities, defined as
\begin{align}
	\tides = \frac{2}{3} k_2 \left(\frac{R}{m}\right)^5
\end{align}
where $k_2$ is the gravitational Love number (which depends on the EoS), $R$ is the NS radius, and $m$ is the mass.
While the precise functional form of $\Lambda(m)$ can vary, the general trend is that a larger mass leads to higher compactness and a significantly decreased tidal response.
See Fig.~\ref{fig:ns_tides} for an example.

We expect $\Lambda = 0$ for non-spinning BHs (see Refs.~\cite{fangTidalCouplingSchwarzschild2005,damourRelativisticTidalProperties2009,binningtonRelativisticTheoryTidal2009}), but a consensus has not yet been reached for spinning BHs.
Spinning BHs could either have $\Lambda = 0$ \cite{landryTidalDeformationSlowly2015,goldbergerNonconservativeEffectsSpinning2021} or $\Lambda \sim O(10^{-3})$ \cite{letiecTidalLoveNumbers2021,letiecSpinningBlackHoles2021} depending on the approach used to obtain the tidal Love number (see Refs.~\cite{grallaAmbiguityRelativisticTidal2018,charalambousVanishingLoveNumbers2021}).
However, NSs should have $\Lambda \gtrsim O(10)$ \cite{letiecSpinningBlackHoles2021}.
Because the maximum possible $\Lambda$ for spinning BHs is orders of magnitude smaller than $\Lambda$ for NSs, we model all BHs with $\Lambda = 0$ for simplicity.

In Sec.~\ref{sec:sing_ev}, we discuss estimates for tidal constraints on a single-event basis and point out that our ability to clearly identify the system components is limited to very loud systems with low masses.
This also implies that the likelihood of observing many events with clearly identified components is low.
Sec.~\ref{sec:population} then approaches this problem at the population level and studies whether a large number of events with poorly constrained tidal properties can still be used to learn about the general composition of the population.


\section{Prospects for tidal measurements with individual events}
\label{sec:sing_ev}

We estimate the measurement uncertainty on $\lameff$ for an individual event with the Fisher information matrix~\cite{vallisneriUseAbuseFisher2008}.
In general, this only provides a lower-bound on the measurement uncertainty, but nevertheless allows us to investigate trends with, e.g., component masses and spins.

The Fisher information matrix is defined as an integral over all possible data ($\data$) given some true signal parameterized by $\parvec^\mu$,
\begin{equation}
    \Gamma_{\mu\nu} = - \int d\data \ p(\data|\parvec_\mathrm{true}) \partial_\mu \partial_\nu \ln p(\data|\parvec)
\end{equation}
where $\partial_\mu = \partial / \partial \parvec^\mu$.
In the high signal-to-noise ratio (SNR) limit, a systematic expansion of the likelihood produces a multivariate normal distribution for $\parvec^\mu$ at leading order with precision matrix $\Gamma_{\mu\nu}$ (see, e.g., \citet{vallisneriUseAbuseFisher2008}).
We therefore approximate the likelihood as a Gaussian centered on the maximum-likelihood estimate ($\theta_\mathrm{MLE}^\mu$)
\begin{equation}
    \ln p(\data|\parvec) \sim -\frac{1}{2}(\parvec - \parvec_\mathrm{MLE})^\mu \Gamma_{\mu\nu} (\parvec - \parvec_\mathrm{MLE})^\nu \ .
\end{equation}
and approximate the posterior distribution by assuming a Gaussian prior also centered on $\theta_\mathrm{MLE}^\mu$ with covariance $P_{\mu\nu}$ so that the posterior distribution has covariance
\begin{equation}
    C_{\mu\nu} = (\Gamma + P^{-1})^{-1}_{\mu\nu} \ .
\end{equation}
In the limit $P \rightarrow \infty$, we obtain the result assuming a flat prior on $\theta^\mu$.
Otherwise, the prior can impose limits that roughly approximate physical boundaries.

If the data is a sum of a signal $h(\parvec^\mu)$ and zero-mean, stationary, Gaussian noise described by a one-sided Power Spectral Density (PSD) $S\lb f\rb$, then the Fisher Information Matrix elements simplify to
\begin{equation}
    \Gamma_{\mu\nu} = 4 \int df \frac{\mathcal{R}\{\partial_\mu \tilde{h}^\ast \partial_\nu \tilde{h}\}}{S}
\end{equation}
where $(\cdot)^\ast$ denotes complex conjugation, $\mathcal{R}\{\cdot\}$ denotes the real part of a complex number, and $\tilde{h}\lb f\rb = A\lb f\rb e^{i\psi\lb f\rb}$ (with $A, \psi \in \mathcal{R}$) is the Fourier transform of the signal $h$.
If we further assume that $\theta^\mu$ only affects the signal's Fourier phase, so that
\begin{equation}
    \partial_\mu \tilde{h} = i \tilde{h} \partial_\mu \psi
\end{equation}
then we obtain
\begin{align}
    \Gamma_{\mu\nu}
        & = 4 \int df \left(\frac{A^2}{S}\right) \partial_\mu \psi \  \partial_\nu \psi \nonumber \\
        & = \rho_\mathrm{opt}^2 \left( \frac{\int df (A^2/S) \partial_\mu \psi \partial_\nu \psi}{\int df A^2/S} \right) \label{eqn:fim_psi}
\end{align}
after remembering that the optimal SNR ($\rho_\mathrm{opt}$) is defined as
\begin{equation}
    \rho_\mathrm{opt}^2 = 4 \int df \frac{|\tilde{h}|^2}{S} = 4 \int df \frac{A^2}{S} \ .
\end{equation}
Note that $\Gamma \propto \rho_\mathrm{opt}^2$, and therefore louder signals (higher $\rho_\mathrm{opt}$) will naturally produce more precise measurements.

We adopt a PN waveform model based on \citet{poissonGravitationalWavesInspiraling1995} and include tidal corrections to the phase $\psi\lb f\rb$ taken from Eq. (12) in \citet{flanaganConstrainingNeutronstarTidal2008}.
The full expressions depend on a reference time and phase, the chirp mass ($\chirp = (m_1 m_2)^{3/5}/(m_1+m_2)^{1/5}$), the symmetric mass ratio ($\smr = m_1 m_2 / (m_1+m_2)^2$), two spin parameters ($\beta$ and $\sigma$), and $\lameff$.
We adopt a diagonal prior covariance matrix with finite marginal variances only for the mass ratio and spin parameters ($\sigma_\smr = 0.25$, $\sigma_\beta = 8.5$, and $\sigma_\sigma = 5.0$)~\cite{vallisneriUseAbuseFisher2008, poissonGravitationalWavesInspiraling1995}.
We also adopt the \textsc{Planck18} cosmology to relate detector- and source-frame parameters, although this difference is negligible for low-mass systems with current detectors~\cite{essickWhenSweatSmall2024, Essick:2025zed}.
Finally, we use the power-law approximation to the maximum likelihood EoS from \citet{essickDetailedExaminationAstrophysical2021,essickAstrophysicalConstraintsSymmetry2021} shown in Fig.~\ref{fig:ns_tides}.

\begin{figure*}
    \centering
    \includegraphics[width=\textwidth]{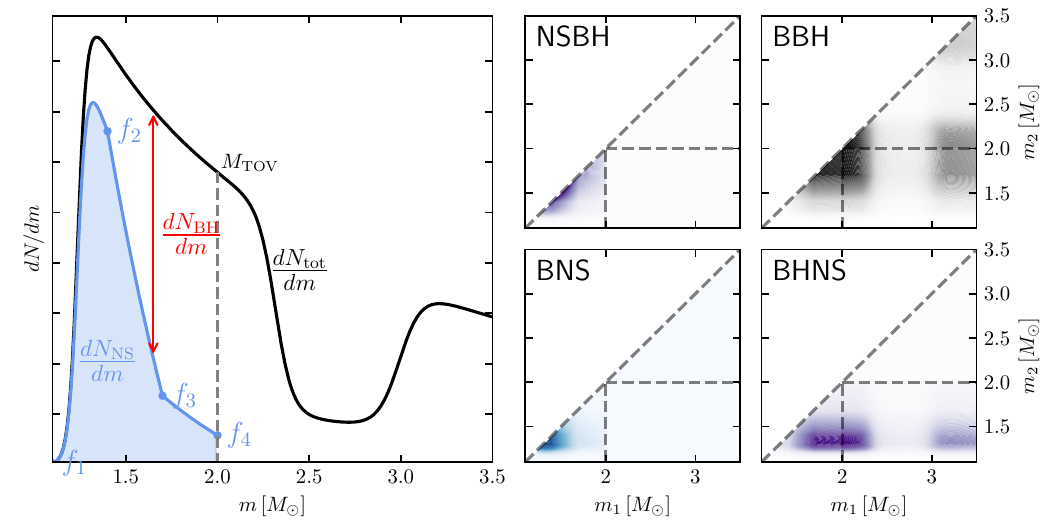}
    \caption{
        Mass distribution assumed in our analysis.
        (\emph{left}) The compact object rate density for (\emph{black}) the overall distribution, (\emph{blue}) NSs, and (\emph{red}) BHs.
        (\emph{right}) The rate density for the four different binary sub-populations in the joint source-frame component mass plane; grey dashed lines denote the boundaries between sub-populations ($M_\mathrm{TOV}$).
    }
    \label{fig:f_NS}
\end{figure*}

Equipped with the Fisher information estimate of the marginal uncertainty in $\lameff$, we examine the dependence on mass in more detail.
Fig.~\ref{fig:fish_tides_fixed_rel_uncert} shows the $\rho_\mathrm{opt}$ needed to reach \result{$\sigma_{\lameff} = \lameff$} and the ratio $\lameff / \sigma_{\lameff}$ at the approximate detection threshold (\result{$\rho_\mathrm{opt} = 10$}; see also~\citet{Essick:2023toz}).
In both cases, only low-mass binaries have a reasonable chance of producing precise constraints on $\lameff$ for realistic $\rho_\mathrm{opt}$.
For high-mass NS and for NS-BH binaries, the ratio of $\lameff/\sigma_{\lameff}$ can be extremely small unless $\rho_\mathrm{opt}$ is extremely large, which will be exceedingly rare (\result{fewer than 1 in $10^{3}$ detected events are expected to have $\rho_\mathrm{opt} > 100$}).
For high-mass NS binaries, this is because both components' tidal deformabilities become small near $M_\mathrm{TOV}$.
For NS-BH binaries, this is because $\lameff$ effectively scales as $\lameff \sim \smr^4 \Lambda_2$, and small $\smr$ can completely overwhelm even very large $\Lambda_2$ for small-mass secondaries.

These estimates provide further quantitative evidence that it may be nearly impossible to distinguish between a NS and a BH near $M_\mathrm{TOV}$ from adiabatic tidal measurements within CBC inspirals.
They also show that it will be easier to precisely measure $\lameff$ for low-mass binaries, and we may even be guaranteed to make a relatively precise measurement ($\lameff/\sigma_{\lameff} \gtrsim 10$) for detected binaries with $m_{1,2} \leq M_\odot$~\cite{Crescimbeni:2024cwh, Golomb:2024mmt}.
Both effects suggest that it may be easier to measure the fraction of NSs in a population at lower masses than at higher masses, which we now confirm with a full hierarchical inference.


\section{Hierarchical Analysis}
\label{sec:population}

We now study whether imprecise measurements of the adiabatic tides can nevertheless be used to identify sub-populations.
We use catalogs of simulated CBCs and conduct hierarchical inference with a suitable population model.

Given a model for the compact object population, described by hyperparameters $\model$ which in turn describe the distribution over single-event parameters $\pnpar$, we model the astrophysical population as an inhomogeneous Poisson process with differential rate-density 
\begin{equation} \label{eq:rate density}
	\frac{d\numdens}{d\pnpar} = \rate \pdf\lb\pnpar\vert\model\rb \ ,
\end{equation}
where $\rate$ is an overall rate and $\pdf\lb\pnpar\vert\model\rb$ is the model for the population.
The joint probability density of detecting $N_{\detec}$ events with data and parameters $\lc\data_i,\pnpar_i\rc$ is then
\begin{widetext}
\begin{equation} \label{eq:hyperpost}
	\pdf( \lc\data_i,\pnpar_i\rc_{i=1}^{N_{\detec}},\model,\rate | N_{\detec} ) = \pdf\lb\model,\rate\rb \rate^{N_{\detec}} e^{-\rate P(\mathbb{D}|\model)} \prod_i^{N_{\detec}} \pdf(\data_i|\pnpar_i) p(\pnpar_i|\model) \ ,
\end{equation}
\end{widetext}
where $\mathbb{D}$ denotes detection, and only detected events appear in the product in Eq.~\ref{eq:hyperpost}.
The selection effects are accounted for in the population-dependent selection function
\begin{align}
    P(\mathbb{D}|\model)
        & = \int d\pnpar \pmf\lb\mathbb{D}\vert\pnpar\rb \pdf\lb\pnpar\vert\model\rb \ .
\end{align}
where $P(\mathbb{D}|\theta)$ is the time-averaged probability that an individual event described by $\theta$ would be detected.
See, e.g., \citet{essickDAGnabbitEnsuringConsistency2023} and~\citet{Essick:2025zed} for more discussion.

Conditioning on the observed data and marginalizing over both the rate (with $p(\mathcal{R}) \propto 1/\mathcal{R}$) and the single-event parameters ($\left\{\theta_i\right\}$) yields a posterior for just population level parameters
\begin{align}
	\pdf\lb\model\vert\left\{\data_i\right\}\rb 
	& \propto \pdf\lb\model\rb \prod_i^{N_{\detec}} \frac{\int d\theta p(\data_i|\theta) p(\theta|\model)}{P(\mathbb{D}|\model)}
\end{align}

\begin{table*}
    \centering
    \caption{
        Fixed population model for component masses, spins, tides, luminosity distance, and orientation.
        Uniform distributions between $X$ and $Y$ are denoted by $U(X, Y)$.
        }
    {\renewcommand{\arraystretch}{1.4}
    \begin{tabular}{@{\extracolsep{0.08cm}} cccccc}
        \hline
        \hline
        parameter                    & description                                     & object type                & \multicolumn{3}{c}{prior} \\
        \hline
        \multirow{12}{*}{$m_{1,2}$}   &  \multirow{12}{*}{source-frame component masses} &  \multirow{12}{*}{BH \& NS} &  \multicolumn{3}{c}{BROKEN POWER LAW + DIP \cite{farahBridgingGapCategorizing2021}} \\
                                     &                                                 &                            &  \multirow{2}{*}{broken power law} & $m_{\mathrm{break}}$    & $2.28 M_\odot$ \\
        \cline{5-6}
                                     &                                                 &                            &                                    & $\alpha_1 = \alpha_2$ 
                        & $-1$\\
        \cline{4-6}
                                     &                                                 &                            &  \multirow{5}{*}{Notch filter}     & $\gamma_{\mathrm{low}}$  & $2.2$ \\
        \cline{5-6}                                     
                                     &                                                 &                            &                                    & $\eta{\mathrm{low}}$     & $100$\\
        \cline{5-6}                                     
                                     &                                                 &                            &                                    & $\gamma_{\mathrm{high}}$ & $3$\\                                     
        \cline{5-6}                                     
                                     &                                                 &                            &                                    & $\eta_{\mathrm{high}}$   & $50$\\                                     
        \cline{5-6}                                     
                                     &                                                 &                            &                                    & $A$                                     
                        & $0.8$\\                                     
        \cline{4-6}                        
                                     &                                                 &                            &  \multirow{2}{*}{High-pass filter} & $m_{\mathrm{min}}$      & $1 M_\odot$ \\
        \cline{5-6}                                     
                                     &                                                 &                            &                                    & $\eta_{\mathrm{min}}$    & $10$\\                                     
        \cline{4-6}                                     
                                     &                                                 &                            &  \multirow{2}{*}{High-pass filter} & $m_{\mathrm{max}}$      & $4 M_\odot$ \\
        \cline{5-6}                                     
                                     &                                                 &                            &                                    & $\eta_{\mathrm{max}}$    & $5.94$\\ 
        \hline
        \multirow{2}{*}{$|\chi_i|$}  & \multirow{2}{*}{spin magnitude}      & BH          & \multicolumn{3}{c}{$U\lb 0,1\rb$} \\
        \cline{3-6}
                                     &                                      & NS          & \multicolumn{3}{c}{$U\lb \chi_{\min}\lb m_i\rb,0.7\rb$, see Eq. \eqref{eqn:chimin}} \\
        \hline
        $\cos(t_i)$                  & spin tilt                            & BH \& NS    & \multicolumn{3}{c}{$U\lb -1,1 \rb$} \\
        \hline
        \multirow{4}{*}{$\tides_i$}  & \multirow{4}{*}{tidal deformability}            & BH                         & \multicolumn{3}{c}{$\tides_i = 0$} \\
        \cline{3-6}
                                     &                                                 & \multirow{4}{*}{NS}        & \multirow{4}{*}{$\tides_i = \tides_0 \left(\frac{m_i}{m_0}\right)^{-\alpha} H(m_i \leq M_\mathrm{TOV})$} & $\tides_0$ & $419.7$\\
        \cline{5-6}                                     
                                     &                                                 &                            &
                   & $m_0$    & $1.366 M_\odot$ \\
        \cline{5-6}                                     
                                     &                                                 &                            & 
                   & $\alpha$ & $5$ \\
        \cline{5-6}
                                     &                                                 &                            & 
                   & $M_\mathrm{TOV}$ & $2\, M_\odot$ \\
        \hline
        $D_L$                        & luminosity distance                 & BH \& NS    & \multicolumn{3}{c}{$\propto dV_c/dD_L$} \\
        \hline
        $\Theta$                     & orientation                         & BH \& NS    & \multicolumn{3}{c}{isotropic/random \cite{dominikDOUBLECOMPACTOBJECTS2015}} \\
        \hline
    \end{tabular}
    }
    \label{tab:model_priors}
\end{table*}

For the inference to proceed, we have to specify a population model (Eq.~\ref{eq:rate density}).
We focus on the mass, spin, and tidal parameters.
Our central assumptions are that the overall compact object distribution is well-known and that it is a mixture of
BHs and NSs:
\begin{align}
    \frac{d \numdens}{d m ds}
	& = \rate \pdf\lb m, s | \model\rb \nonumber\\
	& = \rate \sum_{\object} \pdf\lb m, s\vert\object, \model\rb \pmf\lb\object|\model\rb \ ,
\end{align}
where $\object$ denotes the object type (either BH or NS).
To fully specify the population, we relate the object-dependent distributions $\pdf\lb m, s\vert\object, \model\rb$ to the assumed total distribution, $\pdf\lb m,s|\model\rb$ (or, equivalently, $d\mathcal{N}^{\object}/dmds$ to $d\mathcal{N}/dmds$).
A particularly useful quantity to parameterize the mixture is the ratio of the object-dependent rate-density to the total rate-density, 
\begin{align}
	f_{\object}\lb m, s | \model\rb 
	& = \frac{d \numdens^\object / dmds}{d\numdens / dmds} = \pdf\lb\object\vert m, s, \model\rb \ ,
\end{align}
which corresponds to the fraction of objects at a certain mass (and spin and tidal deformability) that belong to a specific sub-population, which we call the (object) mass fraction.
We study $f_{\ns}\lb m, s | \model \rb$ and obtain $f_{\bh}\lb m, s | \model \rb = 1 - f_{\ns}\lb m, s | \model \rb$.

While we take the shape of $\pdf\lb m, s\vert\object\rb$ as given, we infer the shape of $f_{\ns}\lb m,s\rb$ from the data.
To understand the behavior at a basic level, we further assume $f_\ns$ only depends on $m$ and follows a simple piece-wise constant model.
See the left panel in Fig.~\ref{fig:f_NS}.

Additionally, since we assume a fixed (and known) EoS, we can compute the maximum possible mass of a non-rotating NS ($M_{\mathrm{TOV}}$) and the maximum allowed mass for spinning NS ($M_{\ns}^{\max}$).
See, e.g., \citet{breuMaximumMassMoment2016} and~\citet{mostLowerBoundMaximum2020}.
This fixes the mass fractions in the highest bins in the analysis to $f_{\ns} = 0$ whenever $m > M_{\ns}^{\max}(s)$.

We adopt the parameterized model for the one-dimensional mass distribution from \citet{farahBridgingGapCategorizing2021} and~\citet{farahThingsThatMight2023a, farahTwoKindComparing2024} (Fig.~\ref{fig:f_NS}).
Within the relevant mass range, the distribution is a broken power-law with a notch filter that represents a mass gap.
We convert the one-dimensional mass distribution into a two-dimensional distribution over both component masses with a pairing function~\cite{fishbachPickyPartnersPairing2020,farahTwoKindComparing2024}.
As has become commonplace, we assume the pairing function is a simple power law in the asymmetric mass ratio ($q = m_2 / m_1$) with a preference for equal-mass binaries.
Fig.~\ref{fig:f_NS} provides an example of the two-dimensional distributions for each of the different subpopulations.

We also include a dependence on the spin, as the maximum mass can extend beyond $M_{\mathrm{TOV}}$ for rapidly rotating NSs.
We model this effect following \citet{breuMaximumMassMoment2016} and~\citet{mostLowerBoundMaximum2020}
\begin{widetext}
\begin{equation}
    M_{\ns}^{\max}\lb \spin \vert \chims, M_{\text{TOV}} \rb
        = M_{\text{TOV}} \lb 1+ A_2 \lb \frac{\spin}{\chims}\rb^2 + A_4  \lb \frac{\spin}{\chims}\rb^4\rb
\end{equation}
\end{widetext}
with $A_2 = 0.132$ and $A_4 = 0.0071$.
We assume the Keplerian mass-shedding limit is $\chims=0.7$ \citep{mostLowerBoundMaximum2020}, which is slightly above the bounds inferred from GW170817 in  \citet{essickNonparametricInferenceNeutron2020}, and that the spin magnitudes are uniformly distributed within the allowed range for each mass.
This implies that the population prior is
\begin{widetext}
\begin{align}
    \pdf\lb \spin \vert m, \object=\ns,  \model \rb
        & \propto H(\spin_\mathrm{min} \leq |\spin| \leq \chims) \\
    \spin_{\min}\lb m \rb
        & \equiv
            \begin{cases}
                0 & m \leq M_{\text{TOV}} \\
                \chims \sqrt{ \frac{\sqrt{4 A_4 \frac{m}{M_{\text{TOV}}} +A_2^2-4A_4} -A_2}{2 A_4}} & m > M_{\text{TOV}}
            \end{cases} \ . \label{eqn:chimin}
\end{align} 
\end{widetext}

\begin{table*}
    \centering
    \caption{
        68\% HPD posterior CRs (centered on the mean) for the one- and two-dimensional models for $f_\ns$ based on the hyperposteriors averaged over catalog realizations.
        We consider several catalog sizes and different values of $f_\ns^\mathrm{true}$ (constant throughout the entire NS mass range).
        Some CRs are very asymmetric because the associated distributions have long tails and often peak at the edges of the physically allowable range.
    }
    {\renewcommand{\arraystretch}{1.5}
    \begin{tabular}{@{\extracolsep{0.5cm}} c c c c c c c c}
        \hline
        \hline
        \multirow{2}{*}{$f_\ns^\mathrm{true}$} & \multicolumn{2}{c}{\multirow{2}{*}{parameter}} &  \multicolumn{5}{c}{catalog size} \\
                & & & 10 & 40 & 100 & 160 & 190 \\
    \hline
            \multirow{3}{*}{$0$}
                  & 1-bin & $f_\ns^\mathrm{low}=f_\ns^\mathrm{high}$ & \AFzeroCone & \AFzeroCtwo & \AFzeroCthree & \AFzeroCfour & \AFzeroCfive \\
        \cline{2-8}
                & \multirow{2}{*}{2-bin}
                    & $f_\ns^\mathrm{low}$ & \AGzeroCone & \AGzeroCtwo & \AGzeroCthree & \AGzeroCfour & \AGzeroCfive \\
            \cline{3-8}
                & & $f_\ns^\mathrm{high}$ & \AGoneCone & \AGoneCtwo & \AGoneCthree & \AGoneCfour & \AGoneCfive \\
    \hline
            \multirow{3}{*}{$1/2$}
                  & 1-bin & $f_\ns^\mathrm{low}=f_\ns^\mathrm{high}$ & \BFzeroCone & \BFzeroCtwo & \BFzeroCthree & \BFzeroCfour & \BFzeroCfive \\
        \cline{2-8}
                & \multirow{2}{*}{2-bin}
                    & $f_\ns^\mathrm{low}$ & \BGzeroCone & \BGzeroCtwo & \BGzeroCthree & \BGzeroCfour & \BGzeroCfive \\
            \cline{3-8}
                & & $f_\ns^\mathrm{high}$ & \BGoneCone & \BGoneCtwo & \BGoneCthree & \BGoneCfour & \BGoneCfive \\
        \hline
            \multirow{3}{*}{$1$}
                  & 1-bin & $f_\ns^\mathrm{low}=f_\ns^\mathrm{high}$ & \CFzeroCone & \CFzeroCtwo & \CFzeroCthree & \CFzeroCfour & \CFzeroCfive \\
        \cline{2-8}
                & \multirow{2}{*}{2-bin}
                    & $f_\ns^\mathrm{low}$ & \CGzeroCone & \CGzeroCtwo & \CGzeroCthree & \CGzeroCfour & \CGzeroCfive \\
            \cline{3-8}
                & & $f_\ns^\mathrm{high}$ & \CGoneCone & \CGoneCtwo & \CGoneCthree & \CGoneCfour & \CGoneCfive \\
        \hline
    \end{tabular}
    }
    \label{tab:posteriorCRs}
\end{table*}

For simplicity, we model the tidal deformability as \result{$\tides\lb m\rb = \tides_0 \lb m / m_0\rb^{-\alpha} H(m \leq M_\mathrm{TOV})$ with $\tides_0 = 419.7$, $m_0 = 1.366 M_\odot$, $\alpha = 5$, and $M_\mathrm{TOV} = \result{2 M_\odot}$}, where $H(\cdot)$ represents the indicator function\footnote{The tidal deformability for (rapidly) rotating neutron stars (RNS), which are required for matter partners with $m \in [M_{\mathrm{TOV}},M_{\mathrm{NS}}^{\max}]$, will depend on the compactness of the RNS solutions in this regime, which is expected to be significantly increased in this regime, see e.g. \cite{bejgerParametersRotatingNeutron2013,breuMaximumMassMoment2016}. However, see \cite{salinasAssessingImpactUniform2025} for an example where the compactness can also slightly decrease in going from the maximum-mass stable TOV to the maximum-mass stable RNS. Nevertheless, Fig. \ref{fig:fish_tides_fixed_rel_uncert} indicates that tidal deformabilities for highly compact stars already approaching $M_{\mathrm{TOV}}$ are not expected to be observable and we thus assume them to be zero.}.
See Fig.~\ref{fig:ns_tides}.

We assume standard, uninformative distributions for the rest of the signal parameters, such as isotropic distributions across the sky, isotropic inclinations between the orbital angular momentum and the line-of-sight to the binary, random orbital phase at coalescence and polarization angle, and a uniform-in-volume distance distribution, which corresponds to $p(\rho_\mathrm{opt}) \propto \rho_\mathrm{opt}^{-4}$.
We account for the orientation and sky-position related parameters through the collective parameter $\angproj$ introduced in \citet{finnObservingBinaryInspiral1993}.
Effective distributions for this parameter that lead to realistic measurement uncertainties on the luminosity distance were found in \citet{fishbachDoesBlackHole2018} and \citet{dominikDOUBLECOMPACTOBJECTS2015}.
We use the $3$-parameter fit from \citet{dominikDOUBLECOMPACTOBJECTS2015} to the distribution defined in \citet{finnObservingBinaryInspiral1993} for a one-detector configuration ($n=1$): $\alpha^{(1)} = 1$, $a_2^{(1)} = 0.374222$, $a_4^{(1)} = 2.04216$, and $a_8^{(1)} = -2.63948$.\footnote{Note that our definition of $\angproj$ corresponds to the definition of $w$ in \citet{dominikDOUBLECOMPACTOBJECTS2015}, which differs from the original parameter, $\angproj_{\text{FC}}$, introduced in \citet{finnObservingBinaryInspiral1993}, such that $\angproj = w = \angproj_{\text{FC}} / 4$.}

Tab. \ref{tab:model_priors} summarizes our choices for the population prior.

As part of the hierarchical inference procedure, we mock up reference posterior samples for individual events.
To generate this ``mock PE,'' we follow the same procedure as in \citet{fishbachMostMassiveBinary2020} and~\citet{farahThingsThatMight2023a}.
We make use of a likelihood model that assumes independent (truncated) Gaussian likelihoods for the post-Newtonian detector frame parameters of the GW waveform, $\{ \chirp_z,\smr,\chieff,\chia,\lameff,\dlam,\snr,\angproj\}$.\footnote{This procedure may not perfectly capture the complexity of full parameter estimation, as there likely are correlations between some of these parameters. However, we expect the assumption of independence to be a reasonable model. See, for example, Fig. 12 in~\citet{GW170817-PE} for the weak correlations between $\smr$ and $\lameff$ for GW170817.}
Mock posterior and likelihood distributions are obtained by first scattering the true detector-frame PN parameters of the event $\pnpar^{(\mathrm{true})}$, which are drawn from the population ($\pnpar^{(\mathrm{true})} \sim \pdf\lb\pnpar\vert\model^{(\mathrm{true})}\rb$), according to Gaussians centered on the true parameters, i.e. $\pnpar^{(\mathrm{ML})}\sim \prod_{i=1}^{8} \mathcal{N}\lb\vartheta_i\vert \vartheta_i^{(\mathrm{true})},\sigma_i\rb$.\footnote{$\vartheta_i$ denotes the $i$-th component of the single-event parameter set $\pnpar$ and $\mathcal{N}\lb x \vert \mu,\sigma\rb$ denotes a Gaussian distribution in $x$ with mean $\mu$ and standard deviation $\sigma$.}
These scattered parameters emulate the detector noise and are then used as maximum likelihood values for the  Gaussian likelihoods, such that $\pdf\lb \pnpar\vert\data\rb \approx \prod_{i=1}^{8} \mathcal{N}\lb\vartheta_i\vert \vartheta_i^{(\mathrm{ML})},\sigma_i\rb$.
Mock-posterior samples are then obtained through rejection sampling the prior weighted by the likelihood.
In practice, we use a Gaussian likelihood in $\log\chirp_z$ for the chirp mass and, to account for a change in spread of the likelihoods due to SNR, we always sample the SNR first and then rescale the remaining variances with the SNR, i.e. $\sigma_i \rightarrow \sigma_i / \snr_{\obs}$.
The choices $\{\sigma_{\chirp_z}=0.08,\sigma_{\smr}=0.022,\sigma_\snr=1,\sigma_\angproj=0.21\}$ are as in \citet{fishbachMostMassiveBinary2020} and~\citet{farahThingsThatMight2023a}, which were tuned to match the uncertainties of typical LVK-O3 events.
For the remaining uncertainties, we conduct a similar calibration study based on the events GW170817, GW190425, GW190426, GW190814, GW200105, GW200115~\cite{LIGOScientific:2017vwq, LIGOScientific:2020aai, LIGOScientific:2020zkf, LIGOScientific:2021qlt}, which have component masses consistent with our population model (see Tab. \ref{tab:model_priors}).
We obtain $\{\sigma_{\chieff}=\sigma_{\chia}=0.5,\sigma_{\lameff}=\sigma_{\dlam}=10^4\}$. 


\subsection{results}
\label{sec:hierarchical results}

\begin{figure*}
    \begin{minipage}{0.15\textwidth}
        $f_\ns^\mathrm{true} = 0$ \\
        \vspace{5.5cm}
        $f_\ns^\mathrm{true} = 1/2$ \\
        \vspace{5.5cm}
        $f_\ns^\mathrm{true} = 1$
    \end{minipage}
    \begin{minipage}{0.79\textwidth}
        \includegraphics[trim=0.7cm 1.30cm 0.55cm 0.7cm, clip=True, width=1.0\textwidth]{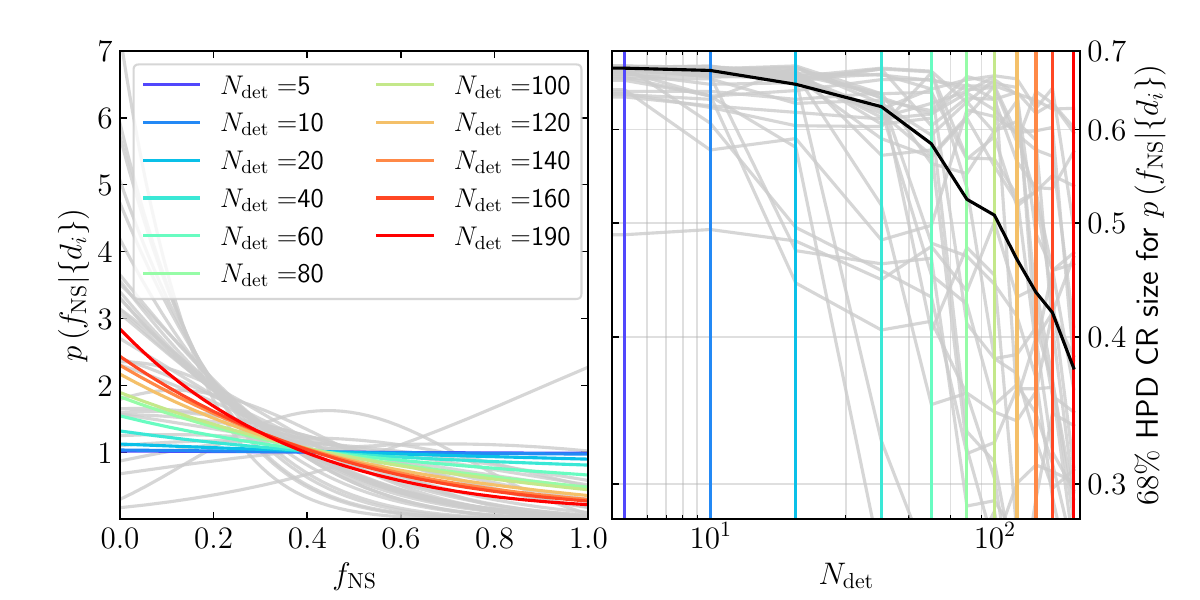} \\
        \includegraphics[trim=0.7cm 1.30cm 0.55cm 0.7cm, clip=True, width=1.0\textwidth]{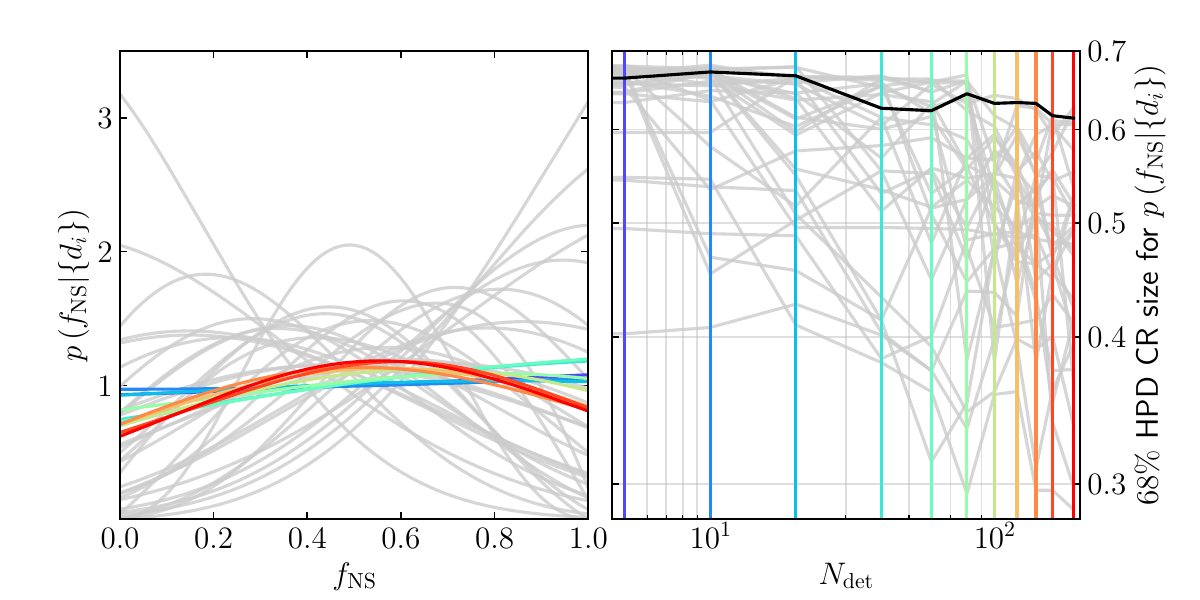} \\
        \includegraphics[trim=0.7cm 0.15cm 0.55cm 0.7cm, clip=True, width=1.0\textwidth]{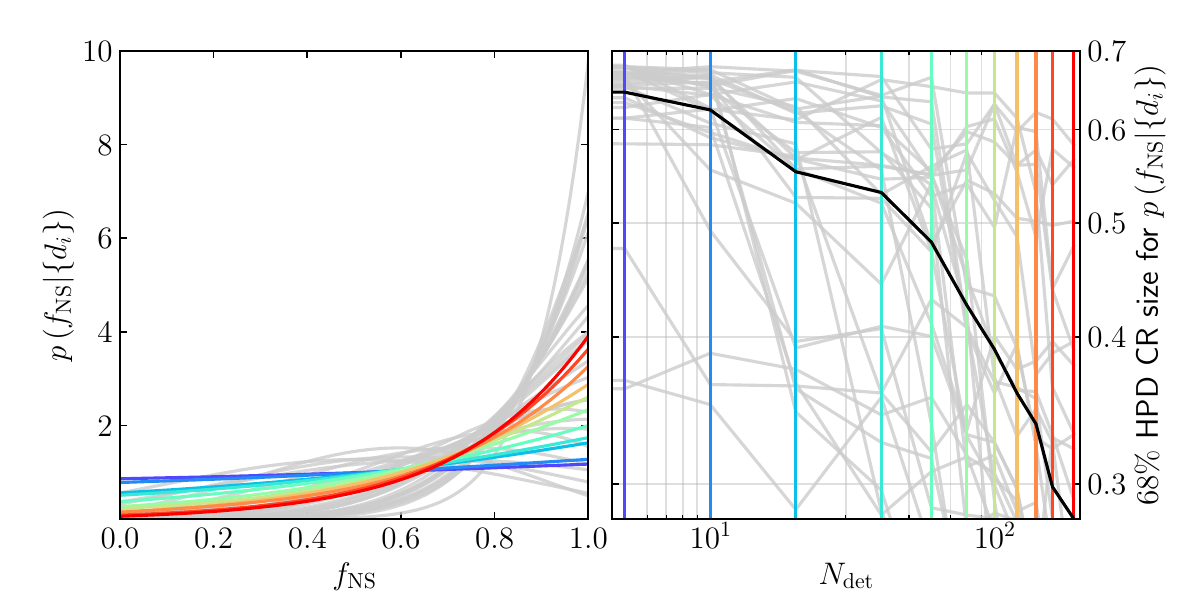}
    \end{minipage}
    \caption{
        (\emph{left}) Hyperposterior for $f_\ns$ when it is assumed to be a constant throughout the entire NS mass range: (\emph{colors}) hyperposterior averaged over catalog realizations at different catalog sizes and (\emph{grey}) hyperposteriors for individual catalog realizations for $N_{\detec} = 190$.
        (\emph{right}) Size of the 68\% HPD credible region vs. catalog size: (\emph{grey}) individual catalog realizations and (\emph{black}) 68\% HPD CR for the hyperposterior averaged over catalog realizations.
        (\emph{colored}) Vertical lines denote the catalog sizes highlighted in the left panels.
    }
    \label{fig:mean_hpd_1d}
\end{figure*}

\begin{figure*}
    \centering
    \includegraphics[trim=0 0.6cm 0 0, clip,width=0.85\textwidth]{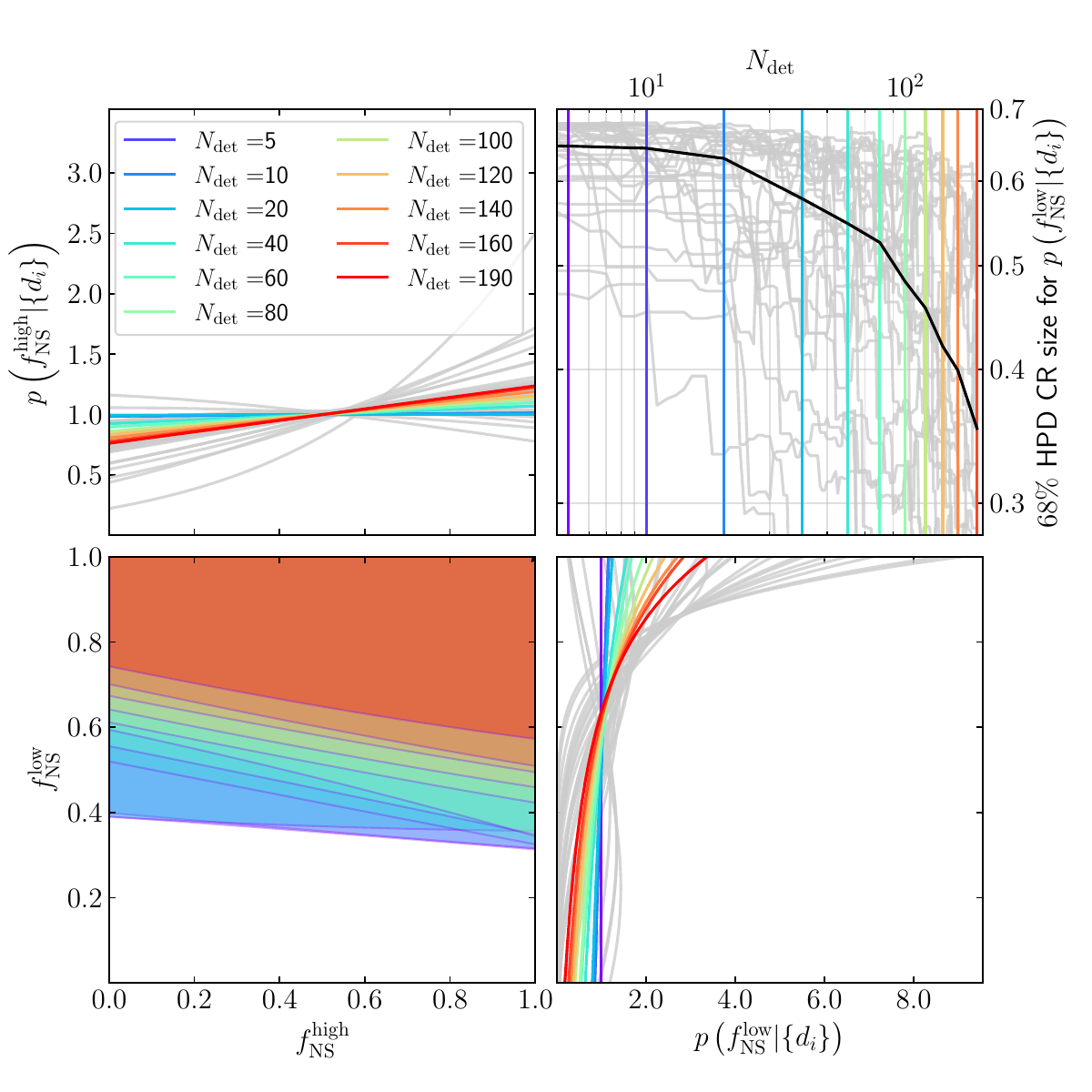}
    \caption{
        Hyperposteriors and credible region sizes for the two-dimensional inference with $f_\ns^\mathrm{true} = 1$.
        Colors match Fig.~\ref{fig:mean_hpd_1d}, and the upper-left and lower-right panels show the marginalized one-dimensional posteriors for $f_\mathrm{NS}$ in the upper ($f_\mathrm{NS}^{\mathrm{high}}$) and lower ($f_\mathrm{NS}^\mathrm{low}$) mass bins, respectively.
    }
    \label{fig:mean_hpd_2d}
\end{figure*}

We study three different true values of $f_\mathrm{NS}$ with two bins within the allowed mass range for NSs.
In principle, the edges of the mass bins that define our piece-wise constant $f_\ns$ should be inferred at the same time as the amplitudes.
However, for simplicity, we fix the upper edge of the upper bin to the maximum possible mass of a NS including the effects of spin (\result{$M_\mathrm{max}^{\mathrm{NS}} = 2.28\,M_\odot$}).
We also assume there are no sub-solar mass objects\footnote{Although searches for such objects are ongoing~\cite{O1-SSM}.} and fix the lower edge of the lower bin to $1\,M_\odot$.
Finally, we place the edge separating these bins at $1.7\,M_\odot$ to approximately separate low-mass systems, for which individual tidal constraints may be informative, from high-mass systems, for which individual tidal measurements are unlikely to be informative.
We also consider BHs up to $4 M_\odot$.
At higher masses, the mass ratio of the binary would lead to a significant suppression of the tidal effects, as discussed in Sec. \ref{sec:sing_ev}.

We generate the three populations, each of which has a constant $f_\mathrm{NS}^\mathrm{true}$ within the entire allowed mass range for NS.
In order to test different astrophysically interesting scenarios, we consider 
\begin{itemize}
    \item no NSs ($f_\ns^\mathrm{true} = 0$),
    \item an equal mixture ($f_\ns^\mathrm{true} = 1/2$), and
    \item only NSs ($f_\ns^\mathrm{true} = 1$).
\end{itemize}
For each model, we generate $5700$ detected events ($\rho_\mathrm{opt} \geq 10$)\footnote{In many cases, selecting events based on their optimal SNR rather than the observed SNR can lead to biases. See, for example, discussion in~\citet{essickDAGnabbitEnsuringConsistency2023}. However, because we assume a fixed overall mass, spin, and distance distribution, and because $f_\mathrm{NS}$ does not affect $P(\mathbb{D}|\lambda)$, the impact of mismodeling the selection effects in this way is negligible.} from which we then generate $30$ catalogs of $190$ events each by randomly drawing events without replacement.

For all of the injected catalogs, we consider two different models and compute hyperposteriors for $f_\ns$.
Specifically, we consider a one-dimensional model where $f_\ns$ is a constant throughout the entire NS mass range (consistent with the injected population) as well as a two-dimensional model in which $f_\ns$ is piece-wise constant within the separate mass bins in the NS mass range.
For each catalog realization, we study the convergence of the hyperposterior with the size of the catalog, and we quantify the precision with the 68\% highest-posterior-density (HPD) credible region (CR).
Figs.~\ref{fig:mean_hpd_1d} and~\ref{fig:mean_hpd_2d} also show the hyperposterior averaged over the different catalog realizations.

In general, and as expected, we find that larger catalogs lead to more precise hyperposteriors.
Generally, we require \result{$O(10)$} events before the hyperposterior no longer appears uniform.
However, even with large catalogs, we are often unable to precisely constrain $f_\ns$.
Table~\ref{tab:posteriorCRs} lists the posterior credible regions for both the one- and two-dimensional models for a few catalog sizes.

For the one-dimensional model, we observe slightly tighter constraints when $f_\ns^\mathrm{true} = 1$ than in the other cases (Tab.~\ref{tab:posteriorCRs} and Fig.~\ref{fig:mean_hpd_1d}).
Injected catalogs of mixed populations ($f_\ns^\mathrm{true} = 1/2$) yield the broadest hyperposteriors, although they are similar in size to what is obtained from catalogs with $f_\ns^\mathrm{true} = 0$.

\begin{figure*}
    \begin{minipage}{0.085\textwidth}
        $f_\ns^\mathrm{true} = 0$ \\
        \vspace{4cm}
        $f_\ns^\mathrm{true} = \frac{1}{2}$ \\
        \vspace{4cm}
        $f_\ns^\mathrm{true} = 1$
    \end{minipage}
    \begin{minipage}{0.9\textwidth}
        \includegraphics[trim=1.7cm 1.6cm 3cm 0.5cm, clip=True, width=1.0\textwidth]{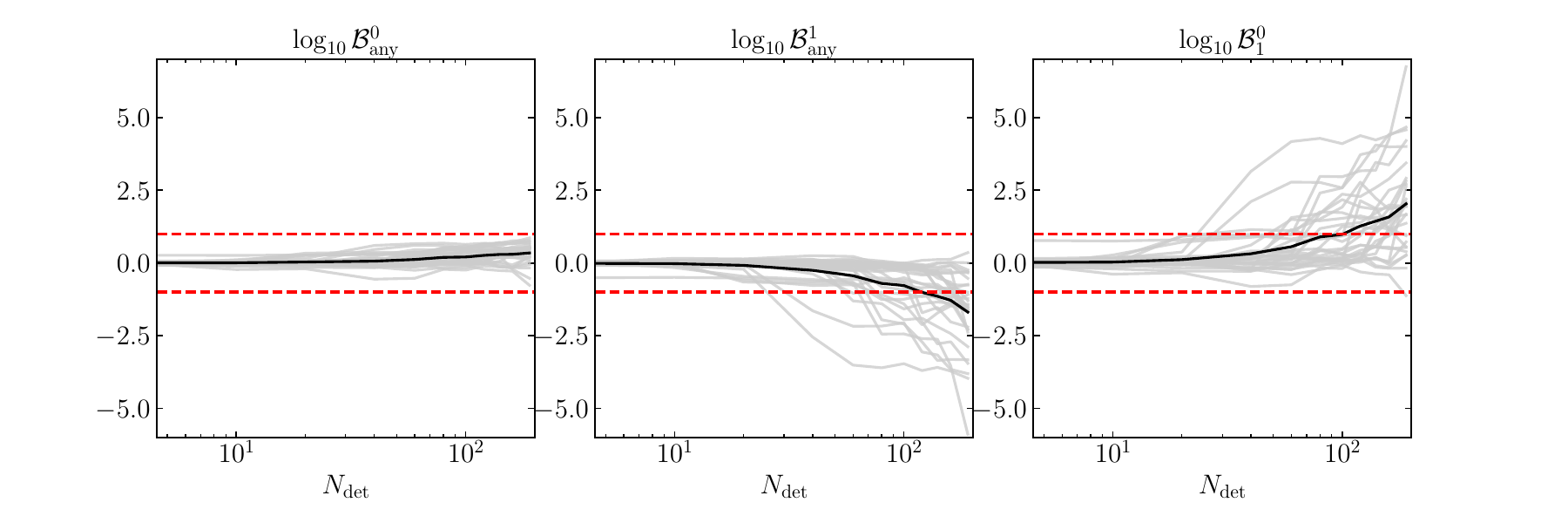} \\
        \includegraphics[trim=1.7cm 1.60cm 3cm 0.5cm, clip=True, width=1.0\textwidth]{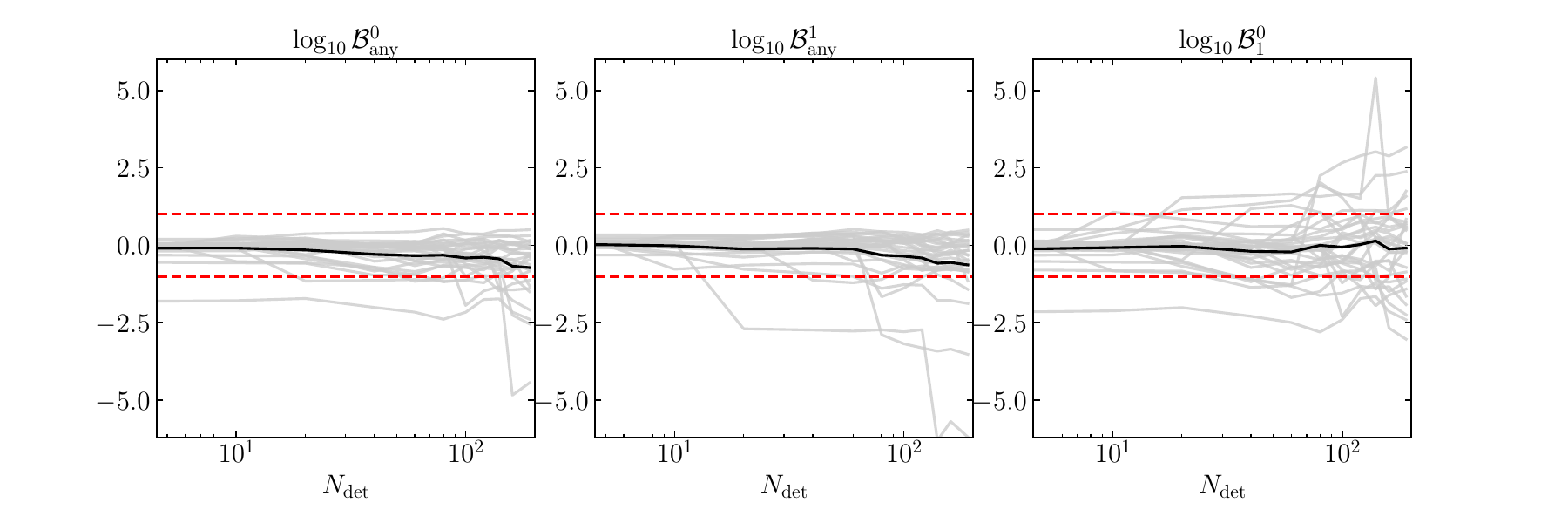} \\
        \includegraphics[trim=1.7cm 0.15cm 3cm 0.5cm, clip=True, width=1.0\textwidth]{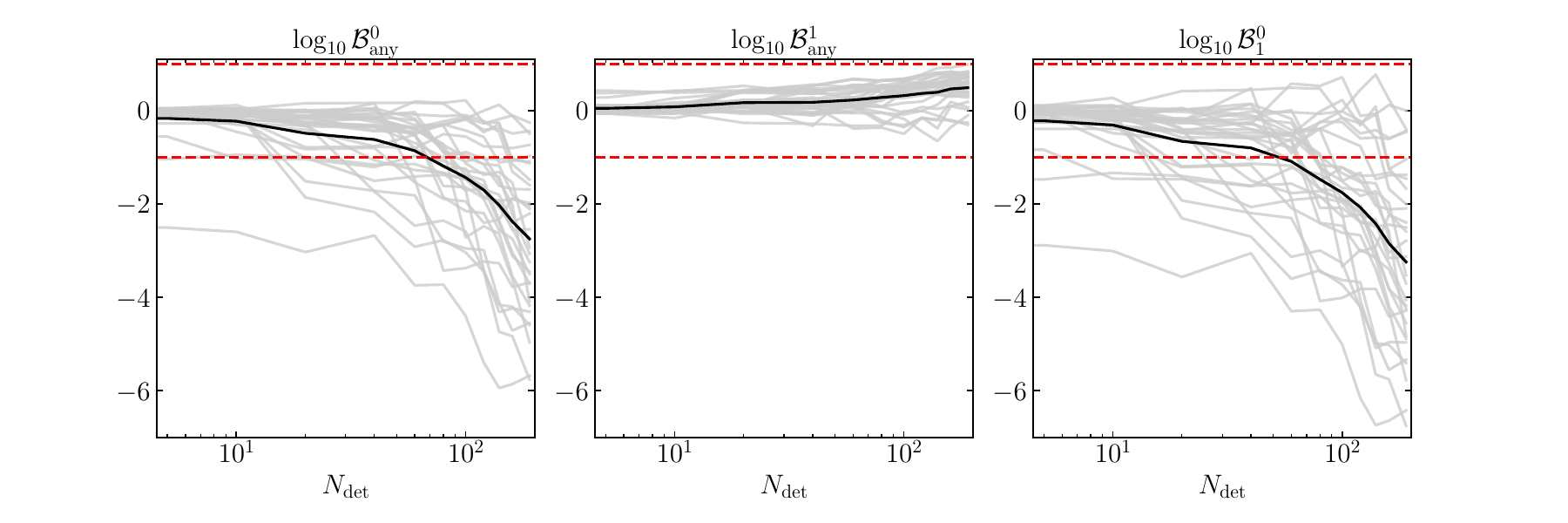}
    \end{minipage}
    \caption{
        (\emph{grey}) Bayes factors ($\mathcal{B}$) as a function of catalog size for the individual catalog realizations.
        (\emph{black} The mean of $\log_{10}\mathcal{B}$ over the different realizations.
        Rows denote (\emph{top to bottom}) $f_\mathrm{NS}^\mathrm{true} = 0$, 1/2, and 1.
        Columns denote (\emph{left to right}) $\log_{10}\mathcal{B}^0_\mathrm{any}$, $\log_{10}\mathcal{B}^1_\mathrm{any}$, and $\log_{10}\mathcal{B}^0_1$.
    }
    \label{fig:SDDR_all_cats}
\end{figure*}

For the two-dimensional inference, we only study the injection campaigns with $f_\ns^\mathrm{true}=1$ in detail, although Table~\ref{tab:posteriorCRs} lists the posterior CRs for all $f_\mathrm{NS}^\mathrm{true}$.
Fig.~\ref{fig:mean_hpd_2d} shows the $68\%$ HPD CRs for the joint hyperposterior over $f_\ns$ in both bins ($f_\ns^\mathrm{low}$ and $f_\ns^\mathrm{high}$).
The two-dimensional posterior confirms our intuition from the Fisher analysis in that less precise tidal information for heavier (more compact) NSs yields a broader posterior for the the upper mass bin.
In fact, the hyperposterior obtained with 190 events still has significant support for all values of $f_\ns^\mathrm{high}$.
In contrast, $f_\ns^\mathrm{low}$ is constrained at a similar level as for the one-dimensional inference.
This shows that most of the information in the one-dimensional constraint comes from low-mass binaries and is only extrapolated to the high-mass systems by the prior assumption that $f_\ns$ is constant within that model.


\subsection{Astrophysical questions}
\label{sec:astro questions}

The outlook for precisely constraining $f_\ns$ with tidal effects during the inspiral of CBCs is generally grim.
Nevertheless, we can still hope to answer astrophysically relevant questions even if the exact behavior of $f_\ns$ is not precisely measured.
In particular, we can try to determine whether BHs and NSs exist at the same mass by trying to rule out $f_\ns = 0$ or $1$ for at least one mass.

We therefore consider the average number of events that it takes for the Bayes factors\footnote{We compute Bayes factors through Savage-Dickey density ratios~\cite{WAGENMAKERS2010158}.}
\begin{equation}
    \mathcal{B}^a_b = \left( \frac{p(f_\ns = a|\left\{\data_i\right\})}{p(f_\ns=b|\left\{\data_i\right\})} \right) \left( \frac{p(f_\ns=b)}{p(f_\ns=a)} \right)
\end{equation}
and
\begin{equation}
    \mathcal{B}^a_\mathrm{any} = \frac{p(f_\ns=a|\left\{\data_i\right\})}{p(f_\ns=a)}
\end{equation}
to either drop below a value of \result{$0.1$} or grow larger than \result{$10$}.
That is, how many events are needed before we can confidently favor a particular value of $f_\ns$ over another (or compared to an agnostic prior) at $90\%$ credibility.

Fig.~\ref{fig:SDDR_all_cats} shows $\log_{10} \mathcal{B}$ for different catalog realizations in grey and the mean over catalog realizations in black. 
In general, we find that even these simpler criteria are difficult to achieve with catalogs of fewer than \result{$100$ events}.
However, when $f_\ns^\mathrm{true} = 1$, we find $\mathcal{B}^0_1 < 0.1$ on average for catalogs of \result{$N_{\detec} \sim O(\result{75})$ events}, meaning we would rule out a population of only BHs with confidence with GW data alone.
That being said, the observation of even a single electromagnetic (EM) counterpart may settle the issue much faster \cite{abbottGravitationalWavesGammaRays2017,abbottMultimessengerObservationsBinary2017}, although EM counterparts have proven to be difficult to find within the large localizations from GW events.


\section{Discussion and Outlook}
\label{sec:discussion}

We investigate the ability of adiabatic tidal signatures from the inspiral of CBCs detected through GWs to determine whether a NS is present within the system.
In agreement with the literature, we find that this is generally a difficult measurement to make for typical SNRs and binary masses.
Interestingly, though, we find that we are almost guaranteed to make a confident detection of tidal signatures for sub-solar mass binaries even at low SNRs near the detection threshold (see also  Refs.~\cite{Crescimbeni:2024cwh, Golomb:2024mmt}).

We then address the question of whether we can learn about the fraction of NSs at a particular mass scale from an entire catalog of CBC detections.
By simulating large catalogs of CBCs and combining them within a hierarchical framework, we find that it will generally be difficult to constrain $f_\ns$ precisely even with catalogs of \result{$O(200)$} events.
As expected, the best constraints are obtained at low masses, and the high-mass fraction (near $M_\mathrm{TOV}$) is only very weakly informed even by the largest catalogs considered.

More optimistically, we find that astrophysically interesting conclusions may still be drawn from more modest catalogs.
In particular, we can expect to confidently rule out populations of only BHs at NS masses with fewer than \result{$O(100)$} events.

These large catalog sizes are likely driven by a few effects.
First, tidal measurements are only likely to be informative for low masses, but GW selection effects favor heavier systems~\cite{Essick:2025zed}.
As such, high-mass BNS and NS-BH detections (which are more likely) will increase the catalog size but provide very little information about $f_\ns$.
We typically find that \result{74.3\%} of our detected catalogs have $m_2 \leq m_1 \leq 1.7\,M_\odot$ (\result{$\lameff / \sigma_{\lameff} \geq 0.32$} at $\rho_\mathrm{opt}=10$) and only \result{15.9\%} have $m_2 \leq m_1 \leq 1.3\, M_\odot$ (\result{$\lameff / \sigma_{\lameff} \geq 1.49$} at $\rho_\mathrm{opt}=10$).

Given the relatively low rate of observed BNS with low masses, it is therefore unlikely that we will reach sufficient catalog sizes to determine $f_\ns$ precisely, or even to answer the simpler question of whether NSs and BHs exist at the same masses, based on GW observations with advanced detectors.
However, next-generation ground-based GW interferometers like Cosmic Explorer and Einstein Telescope are expected to extend the detection horizon for such systems by a factor of $\gtrsim 10$~\cite{essickWhenSweatSmall2024}.
This could increase the detection rate by as much as a factor of $\gtrsim 10^3$.
As such, catalogs of $> O(10^3)$ BNS are not implausible for next-generation detectors, leaving open the possibility of measuring $f_\ns$ directly with GWs.


\acknowledgments

This research was supported in part by Perimeter Institute for Theoretical Physics. Research at Perimeter Institute is supported in part by the Government of Canada through the Department of Innovation, Science and Economic Development and by the Province of Ontario through the Ministry of Colleges and Universities. M.M. acknowledges the support of NSERC, funding reference number No. RGPIN-2019-04684. R.E. is supported by the Natural Sciences \& Engineering Research Council of Canada (NSERC) through a Discovery Grant (RGPIN-2023-03346).

This material is based upon work supported by NSF's LIGO Laboratory which is a major facility fully funded by the National Science Foundation.


\bibliographystyle{aasjournal}
\bibliography{biblio.bib}

\begin{thebibliography}{}
\expandafter\ifx\csname natexlab\endcsname\relax\def\natexlab#1{#1}\fi
\providecommand{\url}[1]{\href{#1}{#1}}
\providecommand{\dodoi}[1]{doi:~\href{http://doi.org/#1}{\nolinkurl{#1}}}
\providecommand{\doeprint}[1]{\href{http://ascl.net/#1}{\nolinkurl{http://ascl.net/#1}}}
\providecommand{\doarXiv}[1]{\href{https://arxiv.org/abs/#1}{\nolinkurl{https://arxiv.org/abs/#1}}}

\bibitem[{Aasi {et~al.}(2015)}]{LIGO}
Aasi, J., {et~al.} 2015, Class. Quant. Grav., 32, 074001,
  \dodoi{10.1088/0264-9381/32/7/074001}

\bibitem[{Abac {et~al.}(2025{\natexlab{a}})}]{GWTC-4}
Abac, A.~G., {et~al.} 2025{\natexlab{a}}.
\newblock \doarXiv{2508.18082}

\bibitem[{Abac {et~al.}(2025{\natexlab{b}})}]{LIGOScientific:2025pvj}
---. 2025{\natexlab{b}}.
\newblock \doarXiv{2508.18083}

\bibitem[{Abbott {et~al.}(2017{\natexlab{a}})Abbott, Abbott, Abbott, \& {et
  al.}}]{abbottGravitationalWavesGammaRays2017}
Abbott, B.~P., Abbott, R., Abbott, T.~D., \& {et al.} 2017{\natexlab{a}}, The
  Astrophysical Journal Letters, 848, L13, \dodoi{10.3847/2041-8213/aa920c}

\bibitem[{Abbott {et~al.}(2017{\natexlab{b}})Abbott, Abbott, Abbott, \& {et
  al.}}]{abbottMultimessengerObservationsBinary2017}
---. 2017{\natexlab{b}}, The Astrophysical Journal Letters, 848, L12,
  \dodoi{10.3847/2041-8213/aa91c9}

\bibitem[{Abbott {et~al.}(2019{\natexlab{a}})Abbott, Abbott, Abbott, \& {et
  al.}}]{abbottGWTC1GravitationalWaveTransient2019}
---. 2019{\natexlab{a}}, Physical Review X, 9, 031040,
  \dodoi{10.1103/PhysRevX.9.031040}

\bibitem[{Abbott {et~al.}(2017{\natexlab{c}})}]{LIGOScientific:2017vwq}
Abbott, B.~P., {et~al.} 2017{\natexlab{c}}, Phys. Rev. Lett., 119, 161101,
  \dodoi{10.1103/PhysRevLett.119.161101}

\bibitem[{Abbott {et~al.}(2018)}]{O1-SSM}
---. 2018, Phys. Rev. Lett., 121, 231103,
  \dodoi{10.1103/PhysRevLett.121.231103}

\bibitem[{Abbott {et~al.}(2019{\natexlab{b}})}]{GW170817-pg}
---. 2019{\natexlab{b}}, Phys. Rev. Lett., 122, 061104,
  \dodoi{10.1103/PhysRevLett.122.061104}

\bibitem[{Abbott {et~al.}(2019{\natexlab{c}})}]{GW170817-remnant}
---. 2019{\natexlab{c}}, Astrophys. J., 875, 160,
  \dodoi{10.3847/1538-4357/ab0f3d}

\bibitem[{Abbott {et~al.}(2019{\natexlab{d}})}]{GWTC-1}
---. 2019{\natexlab{d}}, Phys. Rev. X, 9, 031040,
  \dodoi{10.1103/PhysRevX.9.031040}

\bibitem[{Abbott {et~al.}(2019{\natexlab{e}})}]{GW170817-PE}
---. 2019{\natexlab{e}}, Phys. Rev. X, 9, 011001,
  \dodoi{10.1103/PhysRevX.9.011001}

\bibitem[{Abbott {et~al.}(2020{\natexlab{a}})}]{LIGOScientific:2020aai}
---. 2020{\natexlab{a}}, Astrophys. J. Lett., 892, L3,
  \dodoi{10.3847/2041-8213/ab75f5}

\bibitem[{Abbott {et~al.}(2021{\natexlab{a}})Abbott, Abbott, Abraham, \& {et
  al.}}]{abbottPopulationPropertiesCompact2021}
Abbott, R., Abbott, T.~D., Abraham, S., \& {et al.} 2021{\natexlab{a}}, The
  Astrophysical Journal Letters, 913, L7, \dodoi{10.3847/2041-8213/abe949}

\bibitem[{Abbott {et~al.}(2020{\natexlab{b}})}]{LIGOScientific:2020zkf}
Abbott, R., {et~al.} 2020{\natexlab{b}}, Astrophys. J. Lett., 896, L44,
  \dodoi{10.3847/2041-8213/ab960f}

\bibitem[{Abbott {et~al.}(2021{\natexlab{b}})}]{GWTC-2}
---. 2021{\natexlab{b}}, Phys. Rev. X, 11, 021053,
  \dodoi{10.1103/PhysRevX.11.021053}

\bibitem[{Abbott {et~al.}(2021{\natexlab{c}})}]{LIGOScientific:2021qlt}
---. 2021{\natexlab{c}}, Astrophys. J. Lett., 915, L5,
  \dodoi{10.3847/2041-8213/ac082e}

\bibitem[{Abbott {et~al.}(2023)}]{GWTC-3}
---. 2023, Phys. Rev. X, 13, 041039, \dodoi{10.1103/PhysRevX.13.041039}

\bibitem[{Abbott {et~al.}(2024)}]{GWTC-2.1}
---. 2024, Phys. Rev. D, 109, 022001, \dodoi{10.1103/PhysRevD.109.022001}

\bibitem[{Acernese {et~al.}(2015)}]{Virgo}
Acernese, F., {et~al.} 2015, Class. Quant. Grav., 32, 024001,
  \dodoi{10.1088/0264-9381/32/2/024001}

\bibitem[{Akutsu {et~al.}(2019)}]{KAGRA}
Akutsu, T., {et~al.} 2019, Nature Astron., 3, 35,
  \dodoi{10.1038/s41550-018-0658-y}

\bibitem[{Bejger(2013)}]{bejgerParametersRotatingNeutron2013}
Bejger, M. 2013, Astronomy \& Astrophysics, 552, A59,
  \dodoi{10.1051/0004-6361/201220876}

\bibitem[{Binnington \& Poisson(2009)}]{binningtonRelativisticTheoryTidal2009}
Binnington, T., \& Poisson, E. 2009, Physical Review D, 80, 084018,
  \dodoi{10.1103/PhysRevD.80.084018}

\bibitem[{Branchesi {et~al.}(2023)}]{ET}
Branchesi, M., {et~al.} 2023, JCAP, 07, 068,
  \dodoi{10.1088/1475-7516/2023/07/068}

\bibitem[{Breu \& Rezzolla(2016)}]{breuMaximumMassMoment2016}
Breu, C., \& Rezzolla, L. 2016, Monthly Notices of the Royal Astronomical
  Society, 459, 646, \dodoi{10.1093/mnras/stw575}

\bibitem[{Charalambous {et~al.}(2021)Charalambous, Dubovsky, \&
  Ivanov}]{charalambousVanishingLoveNumbers2021}
Charalambous, P., Dubovsky, S., \& Ivanov, M.~M. 2021, Journal of High Energy
  Physics, 2021, 38, \dodoi{10.1007/JHEP05(2021)038}

\bibitem[{Chen {et~al.}(2020)Chen, {Johnson-McDaniel}, Dietrich, \& {et
  al.}}]{chenDistinguishingHighmassBinary2020}
Chen, A., {Johnson-McDaniel}, N.~K., Dietrich, T., \& {et al.} 2020, Physical
  Review D, 101, 103008, \dodoi{10.1103/PhysRevD.101.103008}

\bibitem[{Chen \& Chatziioannou(2020)}]{chenDistinguishingBinaryNeutron2020}
Chen, H.-Y., \& Chatziioannou, K. 2020, The Astrophysical Journal, 893, L41,
  \dodoi{10.3847/2041-8213/ab86bc}

\bibitem[{Crescimbeni {et~al.}(2024)Crescimbeni, Franciolini, Pani, \&
  Riotto}]{Crescimbeni:2024cwh}
Crescimbeni, F., Franciolini, G., Pani, P., \& Riotto, A. 2024, Phys. Rev. D,
  109, 124063, \dodoi{10.1103/PhysRevD.109.124063}

\bibitem[{Damour \& Nagar(2009)}]{damourRelativisticTidalProperties2009}
Damour, T., \& Nagar, A. 2009, Physical Review D, 80, 084035,
  \dodoi{10.1103/PhysRevD.80.084035}

\bibitem[{Dominik {et~al.}(2015)Dominik, Berti, O'Shaughnessy, \& {et
  al.}}]{dominikDOUBLECOMPACTOBJECTS2015}
Dominik, M., Berti, E., O'Shaughnessy, R., \& {et al.} 2015, The Astrophysical
  Journal, 806, 263, \dodoi{10.1088/0004-637X/806/2/263}

\bibitem[{Essick(2023)}]{Essick:2023toz}
Essick, R. 2023, Phys. Rev. D, 108, 043011, \dodoi{10.1103/PhysRevD.108.043011}

\bibitem[{Essick \& Fishbach(2023)}]{essickDAGnabbitEnsuringConsistency2023}
Essick, R., \& Fishbach, M. 2023, {{DAGnabbit}}! {{Ensuring Consistency}}
  between {{Noise}} and {{Detection}} in {{Hierarchical Bayesian Inference}},
  arXiv, \dodoi{10.48550/arXiv.2310.02017}

\bibitem[{Essick \& Holz(2024)}]{essickWhenSweatSmall2024}
Essick, R., \& Holz, D.~E. 2024, When to {{Sweat}} the {{Small Stuff}}:
  Identifying the Most Informative Events from Ground-Based Gravitational-Wave
  Detectors,  arXiv, \dodoi{10.48550/arXiv.2407.11693}

\bibitem[{Essick {et~al.}(2020)Essick, Landry, \&
  Holz}]{essickNonparametricInferenceNeutron2020}
Essick, R., Landry, P., \& Holz, D.~E. 2020, Physical Review D, 101, 063007,
  \dodoi{10.1103/PhysRevD.101.063007}

\bibitem[{Essick {et~al.}(2021{\natexlab{a}})Essick, Landry, Schwenk, \& {et
  al.}}]{essickDetailedExaminationAstrophysical2021a}
Essick, R., Landry, P., Schwenk, A., \& {et al.} 2021{\natexlab{a}}, Physical
  Review C, 104, 065804, \dodoi{10.1103/PhysRevC.104.065804}

\bibitem[{Essick {et~al.}(2021{\natexlab{b}})Essick, Landry, Schwenk, \& {et
  al.}}]{essickDetailedExaminationAstrophysical2021}
---. 2021{\natexlab{b}}, Physical Review C, 104, 065804,
  \dodoi{10.1103/PhysRevC.104.065804}

\bibitem[{Essick {et~al.}(2021{\natexlab{c}})Essick, Tews, Landry, \& {et
  al.}}]{essickAstrophysicalConstraintsSymmetry2021}
Essick, R., Tews, I., Landry, P., \& {et al.} 2021{\natexlab{c}}, Physical
  Review Letters, 127, 192701, \dodoi{10.1103/PhysRevLett.127.192701}

\bibitem[{Essick {et~al.}(2025)}]{Essick:2025zed}
Essick, R., {et~al.} 2025.
\newblock \doarXiv{2508.10638}

\bibitem[{Evans {et~al.}(2021)}]{CE}
Evans, M., {et~al.} 2021.
\newblock \doarXiv{2109.09882}

\bibitem[{Fang \& Lovelace(2005)}]{fangTidalCouplingSchwarzschild2005}
Fang, H., \& Lovelace, G. 2005, Physical Review D, 72, 124016,
  \dodoi{10.1103/PhysRevD.72.124016}

\bibitem[{Farah {et~al.}(2023)Farah, Edelman, Zevin, \& {et
  al.}}]{farahThingsThatMight2023a}
Farah, A.~M., Edelman, B., Zevin, M., \& {et al.} 2023, The Astrophysical
  Journal, 955, 107, \dodoi{10.3847/1538-4357/aced02}

\bibitem[{Farah {et~al.}(2021)Farah, Fishbach, Essick, \& {et
  al.}}]{farahBridgingGapCategorizing2021}
Farah, A.~M., Fishbach, M., Essick, R., \& {et al.} 2021, arXiv:2111.03498
  [astro-ph].
\newblock \doarXiv{2111.03498}

\bibitem[{Farah {et~al.}(2024)Farah, Fishbach, \&
  Holz}]{farahTwoKindComparing2024}
Farah, A.~M., Fishbach, M., \& Holz, D.~E. 2024, The Astrophysical Journal,
  962, 69, \dodoi{10.3847/1538-4357/ad0558}

\bibitem[{Finn \& Chernoff(1993)}]{finnObservingBinaryInspiral1993}
Finn, L.~S., \& Chernoff, D.~F. 1993, Physical Review D, 47, 2198,
  \dodoi{10.1103/PhysRevD.47.2198}

\bibitem[{Fishbach {et~al.}(2020)Fishbach, Farr, \&
  Holz}]{fishbachMostMassiveBinary2020}
Fishbach, M., Farr, W.~M., \& Holz, D.~E. 2020, The Astrophysical Journal, 891,
  L31, \dodoi{10.3847/2041-8213/ab77c9}

\bibitem[{Fishbach \& Holz(2020)}]{fishbachPickyPartnersPairing2020}
Fishbach, M., \& Holz, D.~E. 2020, The Astrophysical Journal, 891, L27,
  \dodoi{10.3847/2041-8213/ab7247}

\bibitem[{Fishbach {et~al.}(2018)Fishbach, Holz, \&
  Farr}]{fishbachDoesBlackHole2018}
Fishbach, M., Holz, D.~E., \& Farr, W.~M. 2018, The Astrophysical Journal, 863,
  L41, \dodoi{10.3847/2041-8213/aad800}

\bibitem[{Flanagan \&
  Hinderer(2008)}]{flanaganConstrainingNeutronstarTidal2008}
Flanagan, E.~E., \& Hinderer, T. 2008, Physical Review D, 77, 021502,
  \dodoi{10.1103/PhysRevD.77.021502}

\bibitem[{Goldberger {et~al.}(2021)Goldberger, Li, \&
  Rothstein}]{goldbergerNonconservativeEffectsSpinning2021}
Goldberger, W.~D., Li, J., \& Rothstein, I.~Z. 2021, Journal of High Energy
  Physics, 2021, 53, \dodoi{10.1007/JHEP06(2021)053}

\bibitem[{Golomb {et~al.}(2024)Golomb, Legred, Chatziioannou, Abac, \&
  Dietrich}]{Golomb:2024mmt}
Golomb, J., Legred, I., Chatziioannou, K., Abac, A., \& Dietrich, T. 2024,
  Phys. Rev. D, 110, 063014, \dodoi{10.1103/PhysRevD.110.063014}

\bibitem[{Gralla(2018)}]{grallaAmbiguityRelativisticTidal2018}
Gralla, S.~E. 2018, Classical and Quantum Gravity, 35, 085002,
  \dodoi{10.1088/1361-6382/aab186}

\bibitem[{Landry \& Poisson(2015)}]{landryTidalDeformationSlowly2015}
Landry, P., \& Poisson, E. 2015, Physical Review D, 91, 104018,
  \dodoi{10.1103/PhysRevD.91.104018}

\bibitem[{Le~Tiec \& Casals(2021)}]{letiecSpinningBlackHoles2021}
Le~Tiec, A., \& Casals, M. 2021, Physical Review Letters, 126, 131102,
  \dodoi{10.1103/PhysRevLett.126.131102}

\bibitem[{Le~Tiec {et~al.}(2021)Le~Tiec, Casals, \&
  Franzin}]{letiecTidalLoveNumbers2021}
Le~Tiec, A., Casals, M., \& Franzin, E. 2021, Physical Review D, 103, 084021,
  \dodoi{10.1103/PhysRevD.103.084021}

\bibitem[{Most {et~al.}(2020)Most, Papenfort, Weih, \& {et
  al.}}]{mostLowerBoundMaximum2020}
Most, E.~R., Papenfort, L.~J., Weih, L.~R., \& {et al.} 2020, Monthly Notices
  of the Royal Astronomical Society: Letters, 499, L82,
  \dodoi{10.1093/mnrasl/slaa168}

\bibitem[{Poisson \& Will(1995)}]{poissonGravitationalWavesInspiraling1995}
Poisson, E., \& Will, C.~M. 1995, Physical Review D, 52, 848,
  \dodoi{10.1103/PhysRevD.52.848}

\bibitem[{Salinas \& Piekarewicz(2025)}]{salinasAssessingImpactUniform2025}
Salinas, M., \& Piekarewicz, J. 2025, Physical Review C, 112, 015809,
  \dodoi{10.1103/kfsc-8dbd}

\bibitem[{Vallisneri(2008)}]{vallisneriUseAbuseFisher2008}
Vallisneri, M. 2008, Physical Review D, 77, 042001,
  \dodoi{10.1103/PhysRevD.77.042001}

\bibitem[{Wagenmakers {et~al.}(2010)Wagenmakers, Lodewyckx, Kuriyal, \&
  Grasman}]{WAGENMAKERS2010158}
Wagenmakers, E.-J., Lodewyckx, T., Kuriyal, H., \& Grasman, R. 2010, Cognitive
  Psychology, 60, 158, \dodoi{https://doi.org/10.1016/j.cogpsych.2009.12.001}

\end{thebibliography}

\end{document}